\documentclass[12pt,a4paper]{article}
\usepackage{amsmath,amssymb}
\newcommand{\beq}{\begin{eqnarray}}
\newcommand{\eeq}{\end{eqnarray}}

\newcommand{\be}{\begin{equation}}
\newcommand{\ee}{\end{equation}}

\newcommand{\bm}{\begin{multline}}
\newcommand{\fm}{\end{multline}}

\begin{document}
\setlength{\unitlength}{.8mm}

\begin{titlepage} 
\vspace*{0.5cm}
\begin{center}
{\Large\bf Nonlinear integral equations for finite volume excited state energies of the
$O(3)$ and $O(4)$ nonlinear $\sigma$-models}
\end{center}
\vspace{1.5cm}
\begin{center}
{\large \'Arp\'ad Heged\H us}
\end{center}
\bigskip

\begin{center}
Istituto Nazionale di Fisica Nucleare, Sezione di Bologna, \\
       Via Irnerio 46, 40126 Bologna, Italy
\end{center}

\vspace{0.1cm}

\begin{center}
and
\end{center}

\vspace{0.1cm}

\begin{center}
Research Institute for Particle and Nuclear Physics,\\
Hungarian Academy of Sciences,\\
H-1525 Budapest 114, P.O.B. 49, Hungary\\ 
\end{center}
\vspace{2.5cm}
\begin{abstract}
We propose nonlinear integral equations for the finite volume one-particle
energies in the $O(3)$ and $O(4)$ nonlinear $\sigma$-models. The equations
are written in terms of a finite number of components and are therefore
easier to solve numerically than the infinite component excited state
TBA equations proposed earlier. Results of numerical calculations based
on the nonlinear integral equations and the excited state TBA equations
agree within numerical precision.

\end{abstract}

\end{titlepage}



\newsavebox{\Sau}
\sbox{\Sau}{\begin{picture}(140,25) (-70,-12.5)

\put(15,0){\circle{3}}
\put(5,0){\circle{3}}
\put(-5,0){\circle{3}}
\put(-15,0){\circle{3}}
\put(-25,10){\circle{3}}
\put(-25,-10){\circle*{3}}

\put(-3.5,0){\line(1,0){7}}
\put(-13.5,0){\line(1,0){7}}
\put(6.5,0){\line(1,0){7}}

\put(-16.1,1.1){\line(-1,1){7.7}}
\put(-16.1,-1.1){\line(-1,-1){7.7}}

\multiput(17.5,0) (1,0) {6} {\circle*{0.2}}

\put(-21,-11){\makebox(0,0)[t]{{\protect\scriptsize 1}}}
\put(-21,11){\makebox(0,0)[t]{{\protect\scriptsize 2}}}
\put(-12.5,-3){\makebox(0,0)[t]{{\protect\scriptsize 3}}}
\put(-3,-3){\makebox(0,0)[t]{{\protect\scriptsize 4}}}
\put(5.5,-3){\makebox(0,0)[t]{{\protect\scriptsize 5}}}
\put(14,-3){\makebox(0,0)[t]{{\protect\scriptsize 6}}}
\put(28,-2){$b$}
\end{picture}}




\newsavebox{\oharnlie}
\sbox{\oharnlie}{\begin{picture}(140,25) (-70,-12.5)

\put(5,0){\circle{3}}
\put(-5,0){\circle{3}}
\put(-15,0){\circle{3}}
\put(-25,10){\circle{3}}
\put(-25,-10){\circle*{3}}

\put(-3.5,0){\line(1,0){7}}
\put(-13.5,0){\line(1,0){7}}

\put(-16.1,1.1){\line(-1,1){7.7}}
\put(-16.1,-1.1){\line(-1,-1){7.7}}

\multiput(7.5,0) (1,0) {6} {\circle*{0.2}}

\thicklines{
\put(1.5,0){\oval(36.5,15)[]}
}

\put(4,-10){\makebox(0,0)[t]{{\protect\scriptsize  $a(x),\bar{a}(x),F(x)$ }}}

\put(-21,-11){\makebox(0,0)[t]{{\protect\scriptsize 1}}}
\put(-21,11){\makebox(0,0)[t]{{\protect\scriptsize 2}}}
\put(-12.5,-3){\makebox(0,0)[t]{{\protect\scriptsize 3}}}
\put(-3,-3){\makebox(0,0)[t]{{\protect\scriptsize 4}}}
\put(5.5,-3){\makebox(0,0)[t]{{\protect\scriptsize 5}}}
\put(28,-2){$b$}
\end{picture}}



\newsavebox{\onegy}
\sbox{\onegy}{\begin{picture}(80,20) (0,-3.5)

\put(10,0){\circle*{3}}
\put(20,0){\circle{3}}
\put(30,0){\circle{3}}
\put(40,0){\circle{3}}

\put(21.5,0){\line(1,0){7}}
\put(31.5,0){\line(1,0){7}}

\put(16.3,0.4){\line(-1,0){4.8}}
\put(16.3,-0.4){\line(-1,0){4.8}}

\multiput(42.5,0) (1,0) {6} {\circle*{0.2}}

\put(18.5,0){\line(-4,-1){5}}
\put(18.5,0){\line(-4,1){5}}

\put(13,-3){\makebox(0,0)[t]{{\protect\scriptsize 1}}}
\put(23,-3){\makebox(0,0)[t]{{\protect\scriptsize 2}}}
\put(33,-3){\makebox(0,0)[t]{{\protect\scriptsize 3}}}
\put(43,-3){\makebox(0,0)[t]{{\protect\scriptsize 4}}}
\put(57,-2){$a$}
\end{picture}}




\newsavebox{\onegynlie}
\sbox{\onegynlie}{\begin{picture}(80,20) (0,-3.5)

\put(10,0){\circle*{3}}
\put(20,0){\circle{3}}
\put(30,0){\circle{3}}
\put(40,0){\circle{3}}

\put(21.5,0){\line(1,0){7}}
\put(31.5,0){\line(1,0){7}}

\put(16.3,0.4){\line(-1,0){4.8}}
\put(16.3,-0.4){\line(-1,0){4.8}}

\multiput(42.5,0) (1,0) {6} {\circle*{0.2}}

\put(18.5,0){\line(-4,-1){5}}
\put(18.5,0){\line(-4,1){5}}

\thicklines{
\put(36.5,0){\oval(36.5,15)[]}
}

\put(39,-10){\makebox(0,0)[t]{{\protect\scriptsize  $a(x),\bar{a}(x),F(x)$ }}}
\put(13,-3){\makebox(0,0)[t]{{\protect\scriptsize 1}}}
\put(23,-3){\makebox(0,0)[t]{{\protect\scriptsize 2}}}
\put(33,-3){\makebox(0,0)[t]{{\protect\scriptsize 3}}}
\put(43,-3){\makebox(0,0)[t]{{\protect\scriptsize 4}}}
\put(61,-2){$a$}
\end{picture}}



\section{Introduction}

A better theoretical understanding of finite size (FS) effects
is one of the most important problems in Quantum Field Theory (QFT).
The study of FS effects is a useful method of analysing the structure
of QFT models and it is an indispensable tool in the numerical
simulation of lattice field theories. 

Finite size effects can be studied through the volume dependence of the mass gap
of the theory, the usefulness of which is 
demonstrated \cite{1} by the introduction of the L\"uscher-Weisz-Wolff (LWW)
running coupling that enables the interpolation between the
large volume (non-perturbative) and the small volume (perturbative)
regions in both two-dimensional sigma models and QCD.

The study of the L\"uscher-Weisz-Wolff running coupling is useful in the two-dimensional
 $O(N)$ nonlinear sigma (NLS) models, because according to recently performed
high precision Monte Carlo measurements of the LWW running coupling \cite{2}
 the cutoff effects look linear in these models, in contrast to perturbative considerations.
The knowledge of the exact value of the LWW coupling enables one to make better fits
for the cutoff effects, and thus to determine more accurately the functional form of the
lattice artifacts.

Our aim in this paper is to propose nonlinear integral equations (NLIEs)
for the one-particle states of the $O(3)$ and $O(4)$ NLS models that allow for
a fast and accurate numerical calculation of the LWW coupling. Although
this calculation has recently been done in \cite{3} using the excited state
Thermodynamic Bethe Ansatz (TBA) technique \cite{exTBA0,exTBA1,exTBA2,exTBA3,exTBA4,exTBA5},
 the difficulties
corresponding to the infinite number of components in the TBA equations
make the numerical calculations slow and at the same time restrict their
accuracy too. Therefore it is desirable to work with the more convenient NLIE
technique.

Another point of the construction of these excited state NLIEs is to
demonstrate that, similarly to the case of the sine-Gordon model \cite{11a,11},
the NLIE technique can be extended to describe the finite size excited
states also in the family of NLS models.

The infinite set of TBA equations for the ground state of the $O(3)$ and
$O(4)$ NLS models were given in \cite{4a,4b} and \cite{5a,5b} respectively. The derivation
of the equations were based on the fact that NLS models can be represented
as (limits of) certain perturbed conformal field theories \cite{4a,4b,5a,5b}.

The TBA description of the excited states is less systematic in the continuum models 
as it is for the ground state problem. Although a lot of different methods have been
worked out to obtain excited state TBA equations in different models 
\cite{exTBA0,exTBA1,exTBA2,exTBA3,exTBA4,exTBA5}, 
a general construction has not discovered, yet.
The generalization of the TBA equations of the $O(3)$ and $O(4)$ NLS models
 to one-particle excited states was proposed recently \cite{3}. The sigma-model TBA equations consist of
infinitely many components, which makes their numerical analysis difficult.
 
 The NLIEs for the ground state of the $O(3)$ NLS model were proposed in \cite{7}, based on the
 statement, according to which the $O(3)$ NLS model can be expessed as a certain limit
 of appropriately perturbed $Z_{N}$ parafermion conformal field theories \cite{4a}.
 
 The ground state NLIEs for the $O(4)$ NLS model were derived in \cite{8}, using the integrable lattice
 regularization of the model \cite{9a,9b,9c}.
 
 Our main purpose in this paper to propose excited state NLIEs for the one-particle states
 of the $O(3)$ and $O(4)$ NLS models. This is achieved in sections 4 and 5 using the assumption
 that the excited state NLIEs differ from the ground state ones only in additional source terms, and
 if it is necessary in additional quantization conditions.
 
 The paper is organised as follows. In section 2 we recall the TBA integral equations, Y-systems 
 and NLIEs corresponding to the ground state problem. In section 3 we briefly summarize the
 one-particle TBA equations of the models. In sections 4 and 5 we propose excited state NLIEs
 for the $O(4)$ and $O(3)$ NLS models respectively. Numerical solutions of the NLIEs and their
 comparison to perturbation theory and to TBA results are discussed in section 6.
 Finally our conclusions are summarized in section 7.

\section{The ground state problem of the $O(4)$ and $O(3)$ NLS models (TBA and NLIE)}

In this section we give a short review of the ground state TBA equations, Y-systems, and
nonlinear integral equations (NLIEs) for the $O(4)$ and $O(3)$ NLS models.

The ground state energy of a two-dimensional integrable model enclosed in a finite box with periodic 
boundary conditions can be determined by the solutions of the TBA integral equations \cite{Zam}.
The TBA equations of the $O(4)$ and $O(3)$ NLS models can be encoded in infinite Dynkin-diagrams.
(See fig. $1a$ and $1b$)

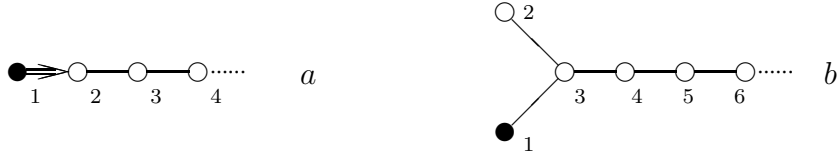
\begin{figure}[htbp]
\begin{center}
\begin{picture}(280,30)(-140,-15)
\put(-120,2) {\usebox{\onegy}}
\put(-74,-7) {\usebox{\Sau}}
\put(-130,-15){\parbox{130mm}{\caption{ \label{6f}\protect {\small
Dynkin-diagrams associated with the Y-systems of the $a$.) $O(4)$ and $b$.) $O(3)$ $\sigma$-models. }}}}
\end{picture}
\end{center}
\end{figure}

 The unknown functions $y_a$ are associated to nodes of the Dynkin-diagram
and the TBA equations are of the form \cite{4a,4b,5a,5b}
\begin{equation} \label{1}
y_a(x)=\exp \left \{ \delta_{a1} \mathcal{D}(x) +
 \sum\limits_{b=1}^{\infty} I_{ab} \ (K*\log Y_b)(x) \right\}, \qquad a=1,2,... 
\end{equation}
where
\begin{equation}
Y_a(x)=1+y_a(x), \qquad \ \ K(x)=\frac{1}{4 \cosh \frac{\pi}{2}x }, 
\qquad \ \  \mathcal{D}(x)=-ml \cosh \left( \frac{\pi}{2} x \right), \label{K}
\end{equation}
the $*$ denotes the convolution, i.e. $(f*g)(x)=\int\limits_{-\infty}^{\infty} dy \ f(x-y)g(y)$, 
$m$ is the mass gap in infinite volume, $l$ is the box size 
and $I_{ab}$ is the incidence matrix of the Dynkin-diagram.
The TBA equations of the $O(4)$ NLS model correspond to the diagram shown in figure $1a$ \cite{5a,5b},
where the oriented double line at the beginning of the diagram means 
\begin{equation}
I_{12}=2, \qquad \qquad I_{21}=1.
\end{equation}  
The TBA equations of the $O(3)$ NLS model are encoded into a $D_{\infty}$ diagram shown in figure
\ref{6f}$b$. The ground state energy can be calculated from the solutions of the TBA equations
\cite{4a,4b,5a,5b} 
\begin{equation}
E_0(l)=-\frac{m}{4} \int\limits_{-\infty}^{\infty} dx \ \cosh \left(\frac{\pi x}{2}  \right) \  
\log Y_1(x). \label{e0}
\end{equation}
The solutions of the TBA equations also satisfy the so called Y-system equations \cite{16}
\begin{equation}
y_a(x+i) \ y_a(x-i)=\prod_{b} {Y_b (x)}^{I_{ab}}.  \label{Y}
\end{equation}

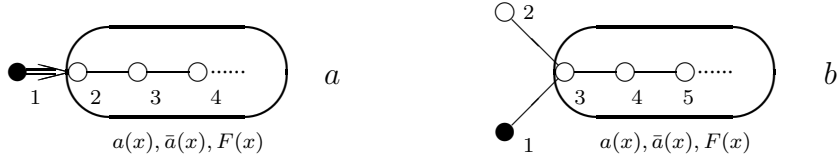
\begin{figure}[htbp]
\begin{center}
\begin{picture}(280,30)(-140,-15)
\put(-120,2) {\usebox{\onegynlie}}
\put(-74,-7) {\usebox{\oharnlie}}
\put(-130,-20){\parbox{130mm}{\caption{ \label{6g}\protect {\small
Graphical representation for the NLIEs of the $a$.) $O(4)$ and $b$.) $O(3)$ $\sigma$-models. }}}}
\end{picture}
\end{center}
\end{figure}
\vspace{0.5cm}
 There is another alternative to calculate the ground state energy of the $O(4)$
 NLS model, namely the nonlinear integral equation technique. The nonlinear integral 
 equations for the ground state of the $O(4)$ NLS were derived in \cite{8} using the light-cone
 lattice approach of the model \cite{9a}. The nonlinear integral equations in this case are of the form
 \begin{eqnarray}
\log y_{1}(x) &=& \mathcal{D}(x)+2 (K^{+\gamma}*\log U)(x)+
                                         2(K^{-\gamma}*\log \bar{U})(x), \nonumber  \\
\log a(x) &=&  (F*\log  U)(x)- (F^{+2(1-\gamma)}*\log \bar{U} )(x)  \nonumber \\
               &+&  (K^{-\gamma}*\log Y_{1})(x), \label{6}   \\
\log \bar{a}(x) &=&  (F*\log  \bar{U})(x)- (F^{-2(1-\gamma)}*\log  U )(x) \nonumber \\  
                          &+&  (K^{+\gamma}*\log Y_{1})(x), \nonumber \\
 U(x)&=&1+a(x), \quad \bar{U}(x)=1+\bar{a}(x), \quad Y_1(x)=1+y_1(x), \nonumber 
\end{eqnarray}
where $0<\gamma<1/2$ is an arbitrary fixed parameter, 
\begin{equation}
F(x)=\int\limits_{-\infty}^{\infty} \frac{dq}{2 \pi} \  \frac{e^{-|q|-iqx}}{2 \cosh(q) }, \label{F}
\end{equation}
and we have used the notation
\begin{equation}
f^{\pm \eta}(x)=f(x \pm i \eta).
\end{equation}
In this case the form of the ground state energy is the same as it is in the TBA case (\ref{e0}).
 Equations (\ref{6}) contain only three real unknown functions because $y_1(x)$ and $Y_1(x)$ are real and $\bar{a}(x)$ 
 is the complex  conjugate of  $a(x)$, therefore they serve an efficient basis for numerical calculations.  
 The graphical notation of these equations is depicted in figure \ref{6g}$a$.
 The big \lq\lq bubble" denotes the complex auxiliary functions which resum the contributions of those
TBA nodes, which are inside it. In our notation the names of the complex unknown
functions and the kernel function are indicated.
  
  In \cite{8} the eqs. (\ref{6}) were derived on a Bethe Ansatz solvable lattice and certainly the TBA equations 
  of the model can be also
  derived from the Bethe Ansatz solution of the model. Thus, it turns out that the function $y_1(x)$ of eqs.
  (\ref{6}) are exactly the same as the function $y_1(x)$ of the TBA equations (\ref{1}). Furthermore the connection
  between the complex $a(x)$ and $\bar{a}(x)$ variables and the TBA variables can be expressed by the following formula
  \begin{equation}
  U(x+i \gamma)\ \bar{U} (x-i \gamma)=Y_2(x), \label{9}
  \end{equation}
which is a very important formula from the NLIE technique point of view, because this relation allows one
to reduce the infinite component TBA system to a finite component NLIEs.  



NLIEs are also available for the ground state of the $O(3)$ NLS model. In \cite{7}
NLIEs were proposed to a class of perturbed parafermion conformal field
theories, which reduce to the $O(3)$ NLS model in a certain limit \cite{4a}.
Taking the appropriate limit the NLIEs take the form
 \begin{eqnarray}
\log y_{1}(x) &=& \mathcal{D}(x)+\log y_{2}(x), \nonumber \\
\log y_{2}(x) &=& (K^{+\gamma}*\log U)(x)+(K^{-\gamma}*\log \bar{U})(x), \nonumber  \\
\log a(x) &=&  (F*\log  U)(x)- (F^{+2(1-\gamma)}*\log \bar{U} )(x)  \nonumber \\
               &+&  (K^{-\gamma}*\log Y_{1})(x)+(K^{-\gamma}*\log Y_{2})(x), \label{10}   \\
\log \bar{a}(x) &=&  (F*\log  \bar{U})(x)- (F^{-2(1-\gamma)}*\log  U )(x) \nonumber \\  
                          &+&  (K^{+\gamma}*\log Y_{1})(x)+(K^{+\gamma}*\log Y_{2})(x), \nonumber \\
 U(x)&=&1+a(x), \ \ \ \bar{U}(x)=1+\bar{a}(x), \ \ \ Y_a(x)=1+y_a(x), \ \ \ a=1,2 \nonumber 
 \end{eqnarray}
where $0<\gamma<1/2$ is an arbitrary fixed parameter, the kernel and source functions are the same as in
(\ref{K}),(\ref{F}) and the ground state energy is given by the formula (\ref{e0}) as in the TBA case.
The graphical representation of these equations is given in figure \ref{6g}$b$.

Although these equations are only conjectured ones, one can recognise that the functions $y_1(x)$
and $y_2(x)$ are the same as the ones of the corresponding TBA equations (\ref{1}), 
$a(x)$ and $\bar{a}(x)$ are the complex conjugates of each other and
the connection between the TBA and NLIE variables can be expressed (similarly to the $O(4)$ case)
by the formula 
\begin{equation}
  U(x+i \gamma)\ \bar{U} (x-i \gamma)=Y_3(x), \label{12}
  \end{equation}
which somehow ensures the reduction of the infinite component Y-system.


\section{TBA equations for the one-particle states of the $O(4)$  and $O(3)$ NLS models}

In this section we briefly review the first excited states TBA equations 
of the $O(4)$ and $O(3)$ NLS models proposed in \cite{3}. 

In \cite{3} the main assumption was (based on previous experience with the
sine-Gordon case) that the Y-system (\ref{Y}) describes not only the ground
state of the model, but remains valid also for the first excited state.
Then L\"uscher's mass gap formula \cite{10}, valid for asymptotically
large volumes, was used to determine the infinite volume solution
of the excited state Y-system, which is sufficient to derive
the one-particle excited state TBA equations. 
It is important from the NLIE technique point of view that 
the infinite volume solutions of the Y-system can be
expressed by an infinite volume t-system, which is of the form
\begin{equation} \label{t}
t_{p}(x+i) \ t_{p}(x-i)=B(x-ip)\ B(x+ip)+t_{p-1}(x) \ t_{p+1}(x), \quad p=1,2,...
\end{equation}
where
\begin{eqnarray}
t_{0}(x) &=& 0, \label{14} \\
B(x) &=& x, \qquad  \quad \mbox{in the} \ \ O(4) \ \ \mbox{case, and }  \label{15} \\ 
\qquad B(x) &=& x^2+1, \qquad \mbox{in the} \ \ O(3) \ \ \mbox{case.} \label{16}
\end{eqnarray}
The solution of (\ref{t}) can be represented in the following form
\begin{equation}
t_{p}(x)=\sum\limits_{l=1}^{p} \lambda_{l}^{(p)}(x), \qquad \lambda_{l}^{(p)}(x)=
B\left[x+i(2l-p-1)\right]. \label{17}
\end{equation} 
Using the infinite volume t-system (\ref{t}), the infinite volume solutions of the
Y-system can be given as
\begin{equation}
y_{p}^{\infty}(x)=\frac{t_{p-1}(x) t_{p+1}(x)}{B(x-ip) B(x+ip)}, \qquad 
Y_{p}^{\infty}(x)=\frac{t_{p}(x-i) t_{p}(x+i)}{B(x-ip) B(x+ip)},  \qquad p=1,2,... \label{18}
\end{equation}
The infinite volume Y-system of the $O(3)$ and $O(4)$ NLS models \cite{3} are of the form
\begin{equation}
y_{p}^{\infty}(x+i) \ y_{p}^{\infty}(x-i)=Y_{p-1}^{\infty}(x) \ Y_{p+1}^{\infty}(x) \qquad
p=2,3,... \label{19}
\end{equation}
where $y_{1}^{\infty}(x)=0$, because the Y-system element corresponding to the massive node
of the Dynkin-diagram tends to zero in the infinite volume limit.
The usefulness of these infinite volume solutions is that, they give the qualitative 
position of the zeroes of the $y_a(x)$ functions of the Y-system, which is sufficient \cite{3,12}
for the derivation of excited state the TBA integral equations. 
Thus the one particle TBA equations of the $O(4)$ NLS model are of the form \cite{3}
\begin{equation}
y_a(x)= {\tau}^2(x) \exp \left \{ \delta_{a1} \mathcal{D}(x) +
 \sum\limits_{b=1}^{\infty} I_{ab} \ (K*\log Y_b)(x) \right\}, \qquad a=1,2,... \label{20}
\end{equation}
where 
\begin{equation} \label{21}
\tau(x)=\tanh\left( \frac{\pi x}{4} \right ),
\end{equation}
and $I_{ab}$ the incidence matrix of figure 1$a$.
The energy of this state takes the form
 \begin{equation} \label{22}
E_1(l)=m-\frac{m}{4} \int\limits_{-\infty}^{\infty} dx  \cosh \left( \frac{\pi x}{2}\right) \log Y_1(x).
\end{equation}

Similarly the one particle TBA equations of the $O(3)$ NLS model can be read off from \cite{3}
and take the form
\begin{eqnarray}
y_{1}(x) &=& e^{\mathcal{D}(x)} \ y_{2}(x), \nonumber \\
y_{2}(x) &=& \tau(x-h_{3}) \ \tau(x+h_{3}) \ \exp \left \{(K*\log Y_3)(x) \right\} \label{23}  \\
y_{a}(x) &=& \tau(x-h_{a-1}) \ \tau(x+h_{a-1}) \ \tau(x-h_{a+1}) \ \tau(x+h_{a+1})  \
e^{ \left \{ \sum\limits_{b=1}^{\infty} I_{ab} \ (K*\log Y_b)(x) \right\} } \nonumber \\ 
\quad a &=& 3,4,... \nonumber 
\end{eqnarray}
and in addition to these integral equations one has to impose quantization conditions
for the $h_a$ zeroes of the $y_a(x)$ functions to ensure that $y_a(\pm h_a \pm i)=-1$,
which is the consequence of the (\ref{Y}) Y-system relations. These quantization equations
can be obtained by the analytic continuation of (\ref{23}) and are of the form
\begin{equation}
\gamma(h_s-h_{s-1})+\gamma(h_s+h_{s-1})+\gamma(h_s-h_{s+1})+\gamma(h_s+h_{s+1}) 
-\frac{1}{i} \ \sum\limits_{b=1}^{\infty} I_{sb} \ (K^{-1}*\log Y_b)(x) = \pi, \label{24} 
\end{equation}
for $s=3,4,...$ where $h_2=0$, $\ \gamma(x)=2 \arctan \tau(x) $ and
\begin{equation}
(K^{-1}*\log Y_b)(x)=i \ \mathcal{P}  \int\limits_{-\infty}^{\infty} \frac{dy}{4} 
\frac{\log Y_b(y)}{\sinh\frac{\pi}{2}(x-y)}, \label{25}
\end{equation}
is a principal value integration.

 Our main goal in this paper to propose NLIEs equivalent to these infinite component
 excited state TBA equations.


\section{Nonlinear integral equations for the first excited state of the $O(4)$ NLS model}

In this section we propose nonlinear integral equations for the first excited state of the $O(4)$
NLS model. The proposal is based on the assumption that the excited states NLIEs differ from the
ground state ones only in additional source terms, plus if it is necessary in quantization conditions. 
In addition we assume that the functional form of these additional source terms 
does not depend directly on the volume. They can depend on the volume only through some
complex objects, on which we impose quantization conditions.
Such objects may be the zeroes of the Y-system elements, or some roots of the 
Bethe Ansatz equations as it happens in the sine-Gordon case \cite{11}.  
 These assumptions are valid for the NLIEs in the
sine-Gordon model, where the form of the excited state NLIEs of ref. \cite{11}
 differ from the ground state NLIE of ref. \cite{11a} only in additional source terms 
 and in quantization conditions.  

In this model one can even give a physical
interpretation of the various terms of the equation. The integral
term represents the contribution of the infinite number of real Bethe Ansatz
roots. The presence of these roots can be regarded as filling the Dirac
sea and thus the presence of this term in the NLIE of the sine-Gordon model is somehow
related to the vacuum structure of the theory. For an excited state
of the model one has to consider a slightly modified configuration
of the Bethe Ansatz roots. It turns out that in the sine-Gordon model
these modifications can be taken into account by adding source terms to
the NLIE of the ground state problem \cite{11}.
Similarly, the excited state TBA equations of the sine-Gordon model
differ from the ground state ones only by some source terms \cite{12}.
Finally, additional support for our assumptions comes from a direct
calculation in the higher-spin vertex model, which reduce to the
$O(4)$ NLS model in the infinite spin limit. These calculations show that
excited state NLIEs differ from the ground state ones again only by
some source terms, which depend on the volume only through some complex
objects, on which quantization conditions have to be imposed.
(As an additional example see \cite{suzuj} for the S=1 special case.)
However, we cannot directly derive the one-particle NLIEs
of the $O(4)$ NLS model from the (limit of the) higher-spin vertex model,
since the solvable lattice model is describing only a subsector of the
Hilbert space of the model \cite{9a,9b} and unfortunately the one-particle states
are not in this subsector.
 
 Accepting these assumptions our task is to find out the necessary source terms of the excited state NLIEs.
 This is achieved similarly to the case of the 
 derivation of the excited state TBA equations of the model,
 where the infinite volume solutions of the equations allowed one to deduce the TBA integral equations \cite{12}.
 The unknown functions of the excited state problem will be denoted in the same way as in eqs. (\ref{6}).
 Assuming that the relation (\ref{9}) holds also for the first excited states,
we can use this relation in the equation for the massive mode in (\ref{1}).
This leads to the following Ansatz for the first excited state NLIE
problem:
 \begin{eqnarray}
\log y_{1}(x) &=& \mathcal{D}(x)+\log \tau^{2}(x)+ 2 (K^{+\gamma}*\log U)(x)+
                                         2(K^{-\gamma}*\log \bar{U})(x), \nonumber  \\
\log a(x) &=& \mathcal{F}_a(x)+(F*\log  U)(x)- (F^{+2(1-\gamma)}*\log \bar{U} )(x)  \nonumber \\
               &+&  (K^{-\gamma}*\log Y_{1})(x), \label{26}   \\
\log \bar{a}(x) &=& \mathcal{F}_{\bar{a}}(x)+ (F*\log  \bar{U})(x)- (F^{-2(1-\gamma)}*\log  U )(x) \nonumber \\  
                          &+&  (K^{+\gamma}*\log Y_{1})(x), \nonumber \\
 U(x)&=&1+a(x), \quad \bar{U}(x)=1+\bar{a}(x), \quad Y_1(x)=1+y_1(x), \nonumber 
\end{eqnarray}
where $0<\gamma<1/2$ is an arbitrary fixed parameter, the energy formula identical to (\ref{22}) 
 and our task is to determine the presently unknown
$\mathcal{F}_a(x)$ and $\mathcal{F}_{\bar{a}}(x)$ source functions. Using the assumption that these
functions do not depend on the volume directly, one can determine them from the infinite volume solution
of the proposed equations, which can be determined from the solution of the infinite volume 
t-system (\ref{t}) and Y-system (\ref{19}). 
Using the construction of \cite{8,13}, the infinite volume solutions of the Ansatz (\ref{26})
 take the form
\begin{equation} 
a_{\infty}(x)=\frac{\lambda_{1}^{(2)}(x+i-i\gamma)}{\lambda_{2}^{(2)}(x+i-i\gamma)}
=\frac{t_{1}(x-i\gamma)}{B\left[x+i(2-\gamma) \right]}=\frac{x-i\gamma}{x+i(2-\gamma)}, \label{27}
\end{equation}
 \begin{equation} \label{28}
\bar{a}_{\infty}(x)=\frac{\lambda_{2}^{(2)}(x-i+i\gamma)}{\lambda_{1}^{(2)}(x-i+i\gamma)}
=\frac{t_{1}(x+i\gamma)}{B\left[x-i(2-\gamma) \right]}=\frac{x+i\gamma}{x-i(2-\gamma)}, 
\end{equation}
\begin{equation} \label{29}
U_{\infty}(x)=1+a_{\infty}(x)=2 \ \frac{x+i(1-\gamma)}{x+i(2-\gamma)},
\end{equation}   
\begin{equation} \label{30}
\bar{U}_{\infty}(x)=1+\bar{a}_{\infty}(x)=2 \ \frac{x-i(1-\gamma)}{x-i(2-\gamma)},
\end{equation}
and $y_{1}^{\infty}(x)=0$. One can check that $a(x)$ is the complex conjugate of
$\bar{a}(x)$ and that these solutions satisfy the relation $U_{\infty}(x+i\gamma) \
\bar{U}_{\infty}(x-i\gamma)=Y_{2}^{\infty}(x)$, which is the infinite volume limit of
(\ref{9}). Once one knows the infinite volume solutions of the Ansatz (\ref{26}), one can substitute
these formulae into them, and can compute explicitly the unknown source functions.
The most convenient way of this is to take the derivative of eqs. (\ref{26}) and work in Fourier space, and
in the end return to the coordinate space and integrate once.
After these simple calculations one gets
\begin{equation} \label{31}
\mathcal{F}_a(x)=\chi(x+i(1-\gamma))+\chi_{K}(x+i(1-\gamma)),
\end{equation}
 \begin{equation} \label{32}
\mathcal{F}_{\bar{a}}(x)=-\chi(x-i(1-\gamma))-\chi_{K}(x-i(1-\gamma)),
\end{equation}
where
\begin{equation} \label{33}
\chi(x)=2\pi i \int\limits_{0}^{x} dy \ F(y)=i \int\limits_{-\infty}^{\infty}
\frac{dq}{q} \ \sin(qx) \ \frac{e^{-|q|}}{2 \cosh(q)},
\end{equation}
\begin{equation} \label{34}
\chi_{K}(x)=2\pi i \int\limits_{0}^{x} dy \ K(y)=i \ \arctan \sinh \left( \frac{\pi x}{2} \right ).
\end{equation}
 Note that these functions are nothing but the odd primitive of the kernel
functions of the integral terms of the ground state equations (\ref{6}), which
is similar to what happens for the NLIE of the sine-Gordon model in the case of
the one hole excitation \cite{11}.
 
We solved these NLIEs numerically  for a number of cases and the results
we found always agreed with those obtained from the numerical solution
of the TBA equations (\ref{20}) (see section 6.). This means that the two methods are equivalent.
On the other hand the TBA equations were tested previously by using
L\"uscher's formula, Monte Carlo measurements \cite{2} and 3-loop perturbation
theory \cite{14}. Thus we are confident that eqs. (\ref{26}) with source terms (\ref{31},\ref{32})
correctly describe the one-particle excited states of the O(4) NLS model.
They are superior to the TBA equations since they contain only three
real unknown functions leading to faster convergence in numerical
calculations.


\section{Excited states nonlinear integral equations for the $O(3)$ NLS model}

In this section following the method described in the previous section we propose
NLIEs for the first excited state of the $O(3)$ NLS model.
According to our assumptions the Ansatz for the NLIEs is of the form
\begin{eqnarray}
\log y_{1}(x) &=& \mathcal{D}(x)+\log y_{2}(x), \nonumber \\
\log y_{2}(x) &=& \log \tau(x-h_3)+\log \tau(x+h_3)+
 (K^{+\gamma}*\log U)(x)+(K^{-\gamma}*\log \bar{U})(x), \nonumber  \\
\log a(x) &=& \mathcal{F}_{{a}}(x)+ (F*\log  U)(x)- (F^{+2(1-\gamma)}*\log \bar{U} )(x)  \nonumber \\
               &+&  (K^{-\gamma}*\log Y_{1})(x)+(K^{-\gamma}*\log Y_{2})(x), \label{35}   \\
\log \bar{a}(x) &=& \mathcal{F}_{\bar{a}}(x)+ (F*\log  \bar{U})(x)- (F^{-2(1-\gamma)}*\log  U )(x) \nonumber \\  
                          &+&  (K^{+\gamma}*\log Y_{1})(x)+(K^{+\gamma}*\log Y_{2})(x), \nonumber \\
 U(x)&=&1+a(x), \ \ \ \bar{U}(x)=1+\bar{a}(x), \ \ \ Y_a(x)=1+y_a(x), \ \ \ a=1,2 \nonumber 
 \end{eqnarray}
where $0<\gamma<1/2$ is an arbitrary fixed parameter, $y_{1}(x)$ and $y_{2}(x)$ are the same variables
as the ones in the excited state TBA eqs. (\ref{23}), and $h_3$ is the zero of the $y_{3}(x)$ Y-system element of
the excited state problem (i.e. $y_{3}(\pm h_3 \pm i)=-1$), therefore an additional quantization equation must be
imposed on this zero. This quantization condition can also be derived from the infinite volume solution of our
equations, and will be discussed at the end of this section. 
The energy expression in turn is the same as the TBA one (\ref{22}).

According to our method first we have to determine the infinite volume solution of our Ansatz (\ref{35}).
This can be achieved using the solutions of the infinite volume t-system (\ref{t}-\ref{17}), and applying the construction
of refs. \cite{8,13}. After some straightforward calculations one gets
\begin{equation} \label{37}
a_{\infty}(x)=\frac{\lambda_{1}^{(3)}(x+i-i\gamma)+\lambda_{2}^{(3)}(x+i-i\gamma)}
{\lambda_{3}^{(3)}(x+i-i\gamma)}
=2 \ \frac{ (x-i\gamma)^2}{\left( x+i(4-\gamma)\right) \cdot \left(x+i(2-\gamma)\right)},
\end{equation}
\begin{equation} \label{38}
\bar{a}_{\infty}(x)=\frac{\lambda_{2}^{(3)}(x-i+i\gamma)+\lambda_{3}^{(3)}(x-i+i\gamma)}
{\lambda_{1}^{(3)}(x-i+i\gamma)}
=2 \ \frac{ (x+i\gamma)^2}{\left( x-i(4-\gamma)\right) \cdot \left(x-i(2-\gamma)\right)},
\end{equation}
\begin{equation} \label{39}
U_{\infty}(x)=1+a_{\infty}(x)
=3 \ \frac{\left( x+h_0+i(1-\gamma) \right) \cdot \left( x-h_0+i(1-\gamma) \right)}
{\left( x+i(4-\gamma) \right) \cdot \left( x+i(2-\gamma) \right) },
\end{equation}
\begin{equation} \label{40}
\bar{U}_{\infty}(x)=1+\bar{a}_{\infty}(x)
=3 \ \frac{\left( x+h_0-i(1-\gamma) \right) \cdot \left( x-h_0-i(1-\gamma) \right)}
{\left( x-i(4-\gamma) \right) \cdot \left( x-i(2-\gamma) \right) },
\end{equation}
\begin{equation} \label{41}
y_{2}^{\infty}(x)=\frac{t_1(x) \ t_3 (x)}{B(x+2i) \ B(x-2i)}
=3 \ \frac{(x+h_0) \ (x-h_0)}{(x+3i) \ (x-3i)},
\end{equation}
\begin{equation} \label{42}
Y_{2}^{\infty}(x)=1+y_{2}^{\infty}(x)=\frac{t_2(x+i) \ t_2 (x-i)}{B(x+2i) \ B(x-2i)}
=3 \ \frac{(x+i) \ (x-i)}{(x+3i) \ (x-3i)},
\end{equation}
where $h_0=\lim\limits_{l \to \infty} h_3=\sqrt{5/3}$, the infinite volume limit of the zero $h_3$, and 
certainly $y_1 (x)=0$ in the infinite volume limit. The infinite volume solutions (\ref{37}-\ref{42}) also satisfy
the important relation $U_{\infty}(x+i\gamma) U_{\infty}(x-i\gamma)=Y_{3}^{\infty}(x)$, which allows one 
to cut the infinite TBA equations.

 Having the infinite volume solutions of our Ansatz,
using the method detailed in the previous section one can calculate the unknown source terms (at least their
infinite volume limit). After simple calculations one gets
\begin{equation} \label{43}
\mathcal{F}_{{a}}(x)=\chi\left(x+h_0 +i(1-\gamma)\right)+\chi\left(x-h_0 +i(1-\gamma)\right)+
2 \chi_{K}\left(x+i(1-\gamma)\right),
\end{equation}
\begin{equation} \label{44}
\mathcal{F}_{\bar{a}}(x)=-\chi\left(x+h_0 -i(1-\gamma)\right)-\chi\left(x-h_0 -i(1-\gamma)\right)-
2 \chi_{K}\left(x-i(1-\gamma)\right),
\end{equation}
according to our assumption these source terms may depend on the volume through the objects, which can be found in
their arguments, namely through the volume dependence of $h_0$. Therefore we replace $h_0$ by its finite volume
value $h_3$ in (\ref{43},\ref{44}), and according to our conjecture
that will be the form of the source functions for finite volume,
 \begin{equation} \label{45}
\mathcal{F}_{{a}}(x)=\chi\left(x+h_3 +i(1-\gamma)\right)+\chi\left(x-h_3 +i(1-\gamma)\right)+
2 \chi_{K}\left(x+i(1-\gamma)\right),
\end{equation}
\begin{equation} \label{46}
\mathcal{F}_{\bar{a}}(x)=-\chi\left(x+h_3 -i(1-\gamma)\right)-\chi\left(x-h_3 -i(1-\gamma)\right)-
2 \chi_{K}\left(x-i(1-\gamma)\right).
\end{equation}
The eqs. (\ref{35}) must be supplemented by the quantization condition of the zero $h_3$, similarly
as it was in the TBA case. The quantization condition can be found out from the infinite volume solutions
(\ref{37}-\ref{42}). One can see from these solutions that $U_{\infty}(\pm h_0-i(1-\gamma))=0$.
Assuming that this relation remains valid also for finite volume, one can infer that 
$U(\pm h_3-i(1-\gamma))=0$, from which follows that 
\begin{equation} \label{47}
a\left( h_3-i(1-\gamma) \right)=-1.
\end{equation}
Taking the logarithm of this, and using eqs. (\ref{35}) with (\ref{45},\ref{46}),
 one gets the quantization condition for $h_3$
\begin{equation} \label{48}
\mathcal{A}(h_3)=\pi,
\end{equation}
where
\begin{eqnarray} \label{49}
\mathcal{A}(x) &=&\frac{1}{i} \ \chi(x-h_3)+\frac{1}{i} \ \chi(x+h_3)+\frac{2}{i} \ \chi_{K}(x)+
\frac{1}{i} \ (K^{-1}*\log Y_1 )(x) \\
&+& \frac{1}{i} \ (K^{-1}*\log Y_2 )(x)+  
\frac{1}{i} \ (F^{-(1-\gamma)}*\log U )(x)- \frac{1}{i} \ (F^{+(1-\gamma)}*\log \bar{U} )(x). \nonumber
\end{eqnarray}
To summarize the eqs. (\ref{35}) with the source functions (\ref{45},\ref{46}), and supplemented by 
the quantization conditions (\ref{48},\ref{49}) make up the one-particle NLIEs of the $O(3)$ NLS
model. In these equations one can also see that the source functions are nothing but the
odd primitives of the kernel functions, as in the $O(4)$ case.
Nevertheless in this case the situation is a bit more complicated, because they contain 
some zeroes in their argument, namely the function $\chi(x)$ contains the zero $h_3$, which  
corresponds to the zero of the $y_3(x)$ (in the $y_3(\pm h_3 \pm i)=-1$ sense), and the
function $\chi_{K}(x)$ contains the zero $h_2=0$, which  
corresponds to the zero of the $y_2(x)$ (in the $y_2(\pm h_2 \pm i)=-1$ sense), but
for this object we do not need to impose quantization equation because of symmetry reasons.

Following from the construction of these equations, one can analytically verify by solving  
iteratively these equations for large $l$ that they give back correctly the same leading order
correction to the infinite volume mass gap as it is predicted by the L\"uscher's formula \cite{10}.     
For not very large volumes we tested these NLIEs through numerical calculations, 
and we found that these equations
give the same numerical results as the earlier proposed excited state TBA equations (\ref{23}-\ref{25})
 (see section 6.).
Because the corresponding TBA system was tested by the results of 3-loop order perturbation theory \cite{14} and 
lattice Monte Carlo measurements \cite{2}, thus our new excited state NLIEs are also consistent with these methods, 
and correctly describe the first excited state energy of the $O(3)$ NLS model.


\section{Numerical results}

 In this section we perform numerical checks on our conjectured one-particle NLIEs.
We solved numerically both the TBA equations and the NLIEs for the ground state and for the one-particle
states at some values of the volume, and we compared the numerical results of the two different methods.
Moreover we solved our conjectured NLIEs in the deep ultraviolet region so as to be able to compare our
numerical results to the predictions of the asymptotically free perturbation theory.

The numerical method used for solving the TBA equations is described in \cite{3}.
The numerical solution of the corresponding NLIEs is rather similar. Namely, we solve numerically 
the NLIEs by iteration. The starting point for the iteration is the $l \to \infty$ solution of the
equations and the procedure converges rapidly. The one-particle state problem for the $O(3)$ NLS model
is more involved since here one step of the iteration includes the calculation of the integrals
occurring in (\ref{35}) together with the calculation of the zero ($h_3$) from the quantization condition (\ref{48},\ref{49}). 
Again, the starting point of the iteration procedure is given by the $l \to \infty$ solution, both for the 
unknown functions and for the position of the zero $h_3$.

We used Simpson's formula, and a cutoff $\Lambda$ for the
numerical evaluations of the integrals running from $- \infty$ to $+ \infty$, in such a way, that 
in the region of $|x|> \Lambda$, the unknown functions are approximated by their infinite volume limit.
The magnitude of the cutoff $\Lambda$ at the values of the volume under consideration has to be 100 
in the $O(4)$ case and 150 in the $O(3)$ case, so that the 9-digit numerical precision could be reached.
Our numerical results are summarized in Tables 1-4. 

\begin{table}
\begin{center}
\begin{tabular}[t]{c||c|c}
$ml$ & $E_0$ (NLIE) & $E_0$ (TBA) \\
\hline
\hline
2 & -0.1620289681(1) & -0.16202897(1) \\
\hline
1 & -0.6437745719(1) & -0.6437746(1) \\
\hline
$10^{-1}$ & -11.273364587(1) & -11.273364(1) \\
\hline
$10^{-2}$ & -127.22634373(1) & -127.2263(1) \\
\hline
$10^{-3}$ & -1343.4090793(1) & -1343.409(1) \\
\hline
$10^{-4}$ & -13865.238816(1) &  \\
\hline
$10^{-5}$ & -141563.8217(1) &  \\
\hline
$10^{-6}$ & -1436683.423(1) &  \\
\hline
\end{tabular}
\end{center}
\caption{{\footnotesize NLIE and TBA results for the ground state energy
in the $O(4)$ NLS model.}}
\end{table}
\begin{table}
\begin{center}
\begin{tabular}[t]{c||c|c}
$ml$ & $E_0$ (NLIE) & $E_0$ (TBA) \\
\hline
\hline
2 & -0.1228466915(1) & -0.1228466(1) \\
\hline
1 & -0.4862495672(1) & -0.4862496(1) \\
\hline
$10^{-1}$ & -8.006985662(1) & -8.006985(1) \\
\hline
$10^{-2}$ & -87.63570019(1) & -87.6357(1) \\
\hline
$10^{-3}$ & -913.9547387(1) & -913.954(1) \\
\hline
$10^{-4}$ & -9374.188294(1) &  \\
\hline
\end{tabular}
\end{center}
\caption{{\footnotesize NLIE and TBA results for the ground state energy
in the $O(3)$ NLS model.}}
\end{table}
\begin{table}
\begin{center}
\begin{tabular}[t]{c||c|c}
$ml$ & $E_1$ (NLIE) & $E_1$ (TBA) \\
\hline
\hline
2 & 0.9923340593(1) & 0.99233406(1) \\
\hline
1 & 0.9383970591(1) & 0.93839706(1) \\
\hline
$10^{-1}$ & -3.004108884(1) & -3.0041089(1) \\
\hline
$10^{-2}$ & -69.83802786(1) & -69.838028(1) \\
\hline
$10^{-3}$ & -901.2815867(1) & -901.28159(1) \\
\hline
$10^{-4}$ & -10260.214298(1) &  \\
\hline
$10^{-5}$ & -111091.0324(1) &  \\
\hline
$10^{-6}$ & -1172575.496(1) &  \\
\hline
\end{tabular}
\end{center}
\caption{{\footnotesize NLIE and TBA results for the one-particle state energy
in the $O(4)$ NLS model.}}
\end{table}
\begin{table}
\begin{center}
\begin{tabular}[t]{c||c|c}
$ml$ & $E_1$ (NLIE) & $E_1$ (TBA) \\
\hline
\hline
2 & 1.02169721(1) & 1.0216972(1) \\
\hline
1 & 1.084208673(1) & 1.084208(1) \\
\hline
$10^{-1}$ & 0.77721084(1) & 0.77718(1) \\
\hline
$10^{-2}$ & -23.6407101(1) & -23.643(1) \\
\hline
$10^{-3}$ & -406.195912(1) & -406.23(1) \\
\hline
$10^{-4}$ & -5150.21619(1) &  \\
\hline
\end{tabular}
\end{center}
\caption{{\footnotesize NLIE and TBA results for the one-particle state energy
in the $O(3)$ NLS model.}}
\end{table}

One can see from these numerical data, that the numerical results served by our NLIEs
agree with those of the TBA equations within the numerical precision.

Now we are able to compare our numerical results to those of the asymptotically free perturbation theory.
Having the numerical values of the ground state energies and the one-particle state energies,
 we can compute numerically the dimensionless finite volume mass gap (LWW coupling)
\begin{equation}
z(ml)=l \cdot [E_1(l)-E_0(l)],
\end{equation}  
for which perturbative results are also available up to 3-loop order \cite{14}.
The perturbative formulae neccesary to compute the 3-loop perturbative mass gap 
can be found in \cite{3}. 
The comparison of our numerical results to the predictions of the perturbation theory 
 can be found in Tables 5 and 6.
\begin{table}
\begin{center}
\begin{tabular}[t]{c||c|c|c}
$ml$ & $z(l)$ (NLIE) & 3-loop PT & 2-loop PT \\
\hline
\hline
$10^{-1}$ & 0.826925570(1) & 0.826130 & 0.8252260 \\
\hline
$10^{-2}$ & 0.573883159(1) & 0.5737662 & 0.5735488 \\
\hline
$10^{-3}$ & 0.442127493(1) & 0.4420969 & 0.4420193 \\
\hline
$10^{-4}$ & 0.360502452(1) & 0.3604916 & 0.3604571 \\
\hline
$10^{-5}$ & 0.30472789(1) & 0.3047233 & 0.3047056 \\
\hline
$10^{-6}$ & 0.26410793(1) & 0.2641057 & 0.2640957 \\
\hline
\end{tabular}
\end{center}
\caption{{\footnotesize NLIE and PT results for the finite volume mass gap $z(ml)$
in the $O(4)$ NLS model.}}
\end{table}
\begin{table}
\begin{center}
\begin{tabular}[t]{c||c|c|c}
$ml$ & $z(l)$ (NLIE) & 3-loop PT & 2-loop PT \\
\hline
\hline
$10^{-1}$ & 0.878419650(1) & 0.876058 & 0.873458 \\
\hline
$10^{-2}$ & 0.639949901(1) &  0.639645 & 0.638874 \\
\hline
$10^{-3}$ & 0.507758827(1) &  0.507669 & 0.507358 \\
\hline
$10^{-4}$ & 0.422397210(1) & 0.422363 & 0.422212 \\
\hline
\end{tabular}
\end{center}
\caption{{\footnotesize NLIE and PT results for the finite volume mass gap $z(ml)$
in the $O(3)$ NLS model.}}
\end{table}
We listed both the 2-loop and 3-loop perturbative results so that one can infer to the accuracy of the
3-loop perturbation theory at the scales under investigation. 
From this comparison, at very small $ml$ values
one can experience a very nice 4-digit agreement in the $O(3)$ case, and an almost 6-digit agreement
in the $O(4)$ case, which is non-trivial, since the conjecture of our NLIEs was based on the
large volume asymptotics of the unknown functions. This perfect agreement makes us confident that
our conjectured one-particle NLIEs describe the exact finite volume one-particle energies in the
$O(3)$ and $O(4)$ NLS models.


\section{Summary and Conclusions}

In this paper we proposed NLIEs for the one-particle states in the $O(3)$ and $O(4)$ NLS models.
The form of these excited state NLIEs are based on the assumption that they
differ from the ground state ones only by some source terms, which may depend on the volume
through some objects in their argument, on which extra quantization conditions must be imposed.
This assumption is mainly motivated by the form of the NLIEs in the
sine-Gordon model 
and by direct calculations
in the higher-spin vertex model (which go to the $O(4)$ NLS model in
the infinite spin limit).  Accepting these assumptions and starting
from the explicit infinite volume solution of the first excited state
Y-system of the models we were able to find the infinite volume limit
of the conjectured equations. This is sufficient to determine the
source terms and the quantization conditions.  It is interesting to note
that in all cases the source terms are the odd primitives of the kernels
occurring in the integral terms of the equations, just as in the case of
the NLIE in the sine-Gordon model.
We have solved these equations numerically and found that results
agree with those obtained previously from numerical solution of the
excited state TBA equations and in the deep ultraviolet regime the
numeical results also agree
with the predictions of the 3-loop perturbation theory.
 This agreement is convincing evidence
for the correctness of our integral equations. The advantage of using
these NLIEs (instead of the infinite set of TBA equations) is that
here the number of unknown function is finite (and small).

An interesting generalization of our results would be to propose excited state NLIEs for
the $\phi_{(id,id,adj)}$ perturbation of the $SU(2)_{L} \times SU(2)_{K} /SU(2)_{L+K}$ models, which were
investigated at rational level K in \cite{7}.

It would also be interesting to extend the NLIE technique for all the excited states
of the $O(3)$ and $O(4)$ NLS models and for
such more complicated TBA systems which can be encoded
into the products of two Dynkin-diagrams \cite{15}.

\vspace{1cm}
{\tt Acknowledgments}

\noindent 
I would like to thank J\'anos Balog for useful discussions and for the critical reading of the manuscript.
The author acknowledges the financial support provided through
the European Community's Human Potential Programme under contract HPRN-CT-2002-00325, 'EUCLID'. 
This investigation was also supported in part by the 
Hungarian National Science Fund OTKA (under T043159) and by INFN Grant TO12.

\end{document}